# Multi-field continuum theory for medium with microscopic rotations

A.A. Vasiliev [a], S.V. Dmitriev [b,c], A.E. Miroshnichenko [d,*]

[a] *Department of Mathematical Modelling, Tver State University, 35 Sadoviy per., 170002 Tver, Russia*
[b] *Institute of Industrial Science, The University of Tokyo, Komaba 4-6-1, Meguro-ku, Tokyo 153-8505, Japan*
[c] *National Institute of Materials Science, 1-2-1, Sengen, Tsukuba, Ibaraki 305-0047, Japan*
[d] *Nonlinear Physics Center, Research School of Physical Sciences and Engineering, The Australian National University, Canberra ACT 0200, Australia*

**Abstract**

We derive the multi-field, micropolar-type continuum theory for the two-dimensional model of crystal having finite-size particles. Continuum theories are usually valid for waves with wavelength much larger than the size of primitive cell of crystal. By comparison of the dispersion relations, it is demonstrated that in contrast to the single-field continuum theory constructed in our previous paper the multi-field generalization is valid not only for long but also for short waves. We show that the multi-field model can be used to describe spatially localized short- and long wavelength distortions. Short-wave external fields of forces and torques can be also naturally taken into account by the multi-field continuum theory.

## 1. Introduction

Continuum approach gives many advantages in physics and mechanics of periodic and aperiodic structures with a great number of structural elements interacting with each other. An important problem of mechanics of generalized continua is to take into account the most essential information about the microscopic structure of matter.

The classical elasticity theory takes into account three degrees of freedom for an infinitesimally small volume, i.e., three components of displacement vector. However, there are many practically important materials having almost rigid atomic clusters with relatively weak interaction between them and one has to consider not only translational but also rotational and perhaps some other degrees of freedom to characterize positions, orientations, and distortions of the particles. Of particular interest are the materials where the framework of rigid clusters is organized in a way that translational and rotational degrees of freedom are coupled (see, e.g., Fig. 1). Some important classes of such materials are the polymorphs of silica ($SiO_2$) where the structural units, $SiO_4$ tetrahedra, shear oxygen atoms in their corners and the energy cost of deformation of tetrahedra is much greater than the cost of their mutual rotations. It has been demonstrated that the rotational degrees of freedom can be responsible for the displacive type of phase transition and incommensurate phase of quartz (Vallade et al., 1992; Wells et al., 2002; Dmitriev et al., 2003a), negative Poisson ratio of cristobalite and quartz (Keskar and Chelikowsky, 1993; Smirnov and Mirgorodsky, 1997; Kimizuka et al., 2000; Alderson and Evans, 2002), and negative thermal expansion of beta-quarts (Smirnov, 1999). Similar effects can be observed in other materials with microscopic rotations, such as perovskites, $SrTiO_3$, containing corner-linked $TiO_6$ octahedra, $KH_2PO_4$ (KDP) family of crystals with comparatively rigid $PO_4$ tetrahedra and others.

Major physical effects related to the materials with rotational degrees of freedom can be discussed theoretically in frame of the rigid-unit-mode (RUM) model (Swainson and Dove, 1993; Dove et al., 1997; Wells et al., 2002). The RUMs are the low-frequency modes in which the clusters can move without distortion. In a real situation, any mode can include components of displacive, rotational and distortive motion, with the relative proportions of the different types varying more or less continuously from low-frequency modes with little distortive component to high-frequency modes with little rigid-unit component (Wells et al., 2002). The origin of auxetic behaviour (i.e., negative Poisson ratio) of materials with finite size particles (molecules) has been discussed in frame of 2D models (Vasiliev et al., 2002; Wojciechowski et al., 2003; Wojciechowski, 2003). Incommensurate structures have been studied in frame of 1D and 2D elastically hinged molecule (EHM) models where the mutual rotations of the finite size particles are very important and all the phonon modes are of nearly RUM type (Dmitriev et al., 1997, 1998, 2000, 2003b; Braun and Kivshar, 2004).

[*] Corresponding author. Tel.: +61-2-6125-9653; fax: +61-2-6125-8588.
   *E-mail address:* andrey.miroshnichenko@anu.edu.au (A.E. Miroshnichenko).

Construction of the continuum theories taking into account microscopic rotations dates back to the work by E. and F. Cosserat (Cosserat and Cosserat, 1909). Cosserat and micropolar continuum theories, in which the field of rotations is incorporated in addition to the displacement fields, have found their applications in many branches of physics and mechanics such as mechanics of granular media (Mühlhaus and Oka, 1996; Suiker et al., 2001), structured materials (Lakes, 1991; Forest et al., 2001), repetitive beam lattice structures (Noor, 1988), physics of dielectric crystals (Pouget et al., 1986; Eringen, 1999; Maugin, 1999; Vasiliev et al., 2002), and others. 2D continuum models for media with microscopic rotations have been developed also in fabric mechanics (Kuwazuru and Yoshikawa, 2004a, 2004b).

Another idea of generalization of the classical continuum theory is taking into account higher order gradient terms (Mühlhaus and Oka, 1996; Fleck and Hutchinson, 1997, 2001; Aifantis, 2003; Peerlings et al., 2001; Askes et al., 2002). A beautiful exposition of some other theories of non-classical material continua may be found in the review by Rogula (1985).

In this paper we construct a *multi-field* micropolar continuum theory for the 2D model of KDP crystal studied in our previous work (Vasiliev et al., 2002). Crystalline solids have translational periodicity and usually the basic idea is to start from a primitive (minimum volume) periodic element of the structure, then define the most important degrees of freedom in it, and finally, to formulate a continuum theory in terms of these degrees of freedom under the assumption that they vary slowly in space. However, for the oriented media, a strong coupling of long and short waves takes place very often and one may need a theory capable of description of both long and short waves. Such a theory can be constructed considering from *more than one* primitive translational cells in an extended periodic cell. To construct a $N$-field continuum theory we consider a macro-cell containing $N$ primitive cells. The number of degrees of freedom in the macro-cell is $N$ times larger than in a primitive cell and consequently, the number of continuum equations of the multi-field theory is $N$ times larger than that for a conventional theory, which will be called single-field theory meaning that there is only one field for each component of vector of generalized displacements. As for the discrete system, consideration of translational cell with more than one primitive cell adds no new physical details. However, continuum analogue constructed for an extended periodic cell with a larger number of fields used to describe each component of displacement gives a possibility to describe the discrete system more accurately (Il'iushina, 1969, 1972; Vasiliev, 1994, 1996).

This paper is organised as follows. In Sec. 2, the 2D discrete model of KDP crystal with finite size particles is described. In Sec. 3 and Sec. 4, we derive the discrete equations of motion for a periodic cell containing two particles; then we derive the corresponding two-field micropolar continuum theory; and, for the sake of comparison, obtain the dispersion relations for the discrete model and continuum theories. In Sec. 5, we derive the four-field micropolar theory and corresponding dispersion relations. In Sec. 6, multi-field modelling of localised distortions in crystal is considered. In Sec. 7 we generalize the obtained results and discuss the difference between the high-gradient and the multi-field continuum theories. Section 8 concludes the paper.

**2. 2D model of crystal with finite size particles**

We consider the 2D microscopic model of KDP crystal offered in Ishibashi and Iwata (2000) and Vasiliev et al. (2002). The model consists of absolutely rigid elastically bound square particles and each particle experiences the action of the rotational background potential (see Fig. 1(a)). The geometry of the model can be described by two parameters, the lattice spacing, $h$, and the parameter

$$A = \frac{a}{\sqrt{2}} \sin \alpha ,  \qquad (1)$$

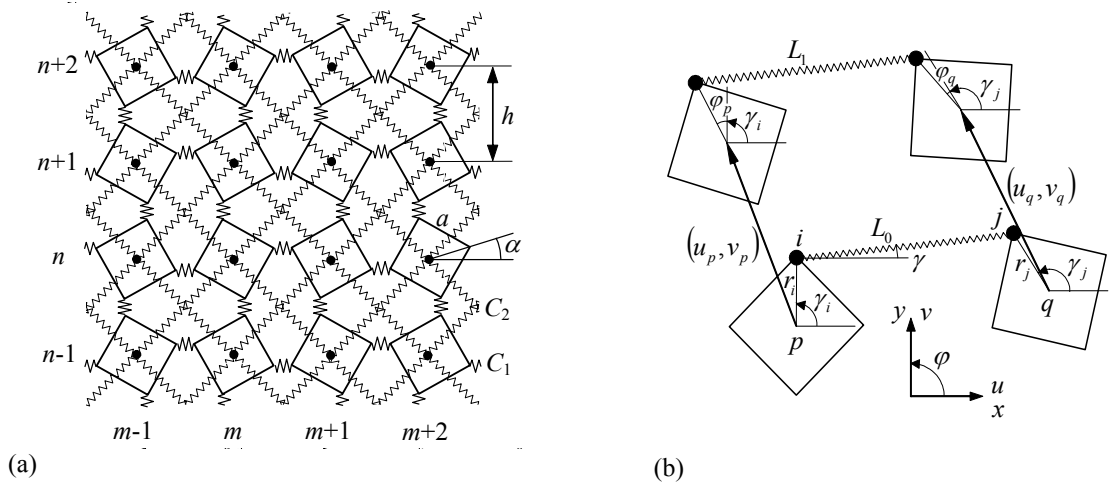

Fig. 1. (a) The 2D microscopic model of a crystal. Absolutely rigid square particles are bound elastically and each particle experiences the action of the rotational background potential. The lattice spacing is $h$, and $a$, $\alpha$ are the size and the orientation angle of particles, respectively. $C_1$ and $C_2$ are the elastic constants for the corner-to-corner and center-to-center bonds, respectively. (b) Definition of displacements $u$, $v$ and $\varphi$, and notations used in calculation of the potential energy of the elastic bonds, Eq. (3).

where $a$ is the size of particles and $\alpha$ is the orientation angle of particles. Elastic bonds with coefficient $C_1$ connect the corners of each particle with the nearest corners of nearest neighbours. Elastic bonds with coefficient $C_2$ connect the centres of particles with the centres of next-nearest neighbours. Each particle has mass $M$, moment of inertia $I$, and experiences the action of the rotational background potential with coefficient $c$. Particles have three degrees of freedom, two components of displacement vector, $u$, $v$, and the angle of rotation $\varphi$.

In order to obtain equations of motion for $p$th particle, the following Lagrangian is constructed,

$$L_p = \frac{1}{2}M\dot{u}_p^2 + \frac{1}{2}M\dot{v}_p^2 + \frac{1}{2}I\dot{\varphi}_p^2 - \sum E_{(p,i)(q,j)} - \frac{1}{2}c\varphi_p^2 + f^x u_p + f^y v_p + f^\varphi \varphi_p, \qquad (2)$$

where $u_p$, $v_p$ are the components of displacements of the center gravity of $p$th particle and $\varphi_p$ is the angle of rotation of this particle; the first three terms give the kinetic energy of the $p$th particle, the fourth term describes the potential energy of eight elastic bonds connecting the $i$th node of $p$th particle with the $j$th node of $q$th particle, the fifth term is the energy of particle in the background potential, and the last three terms give the work of external force with components $f^x$, $f^y$ and external torque $f^\varphi$. Note that the external forces were not taken into account in Vasiliev et al. (2002).

Potential energy of the elastic bond connecting the $i$th node of $p$th particle with the $j$th node of $q$th particle is $E_{(p,i)(q,j)} = (C/2)(L_1 - L_0)^2$, where $C$ is the bond stiffness, $L_0$ and $L_1$ are the bond lengths before and after deformation, respectively. We linearise the change in the bond length, $\Delta L = L_1 - L_0$, with respect to displacements and rotations assuming that they are small and represent the potential energy in the form

$$E_{(p,i)(q,j)} = \frac{1}{2}C[(u_q - u_p)\cos\gamma + (v_q - v_p)\sin\gamma + r_i\varphi_p\sin(\gamma_i - \gamma) - r_j\varphi_q\sin(\gamma_j - \gamma)]^2, \qquad (3)$$

where $C = C_1$ or $C = C_2$ for corner-to-corner and for center-to-center bonds, respectively, and parameters $r_i$, $r_j$, $\gamma$, $\gamma_i$, $\gamma_j$ are related to the undeformed geometry (see Fig. 1(b)). Particularly, $r_i = r_j = a/\sqrt{2}$ and $r_i = r_j = 0$ for corner-to-corner and center-to-center bonds, respectively; $\gamma_i$ is the angle between $x$-axis and the radius vector connecting the center of mass of $p$ th particle with its $i$ th node; $\gamma$ is the angle between $x$-axis and the direction from the node $i$ to the node $j$. Parameters $\gamma_i$, $\gamma_j$ and $\gamma$ can take values $l\pi/2 \pm \alpha$ and $l\pi/4$, respectively, with an integer $l$.

## 3. Discrete model with two particles in a periodic cell

In Vasiliev et al. (2002), we used a unit cell with one particle, $(m,n)$, to write down the three discrete equations of motion and to derive a single-field continuum theory from the discrete equations.

Here we use a cell containing two particles with coordinates $(m,n)$ and $(m+1,n)$ respectively (Fig. 1). In the present case, each particle has subscript index $s$ which is 0 if the sum $m+n$ is even and it is 1 if the sum is odd. There are six degrees of freedom per unit cell, $u_s$, $v_s$, $\varphi_s$, $s=0,1$. We also use this subscript index for forces and torques applied to the particles.

For the two particles in a periodic cell we construct the Lagrangian Eq. (2) and derive six equations of motion. Equations for particles with $s=0$ read

$$M\ddot{u}_0^{m,n} = C_1\left(\Delta_{xx}u_1^{m,n} + 2u_1^{m,n} - 2u_0^{m,n}\right) + \frac{1}{2}C_2\left(\Delta u_0^{m,n} + \Delta_{xy}v_0^{m,n}\right) - AC_1\Delta_x\varphi_1^{m,n} + f_0^x,$$

$$M\ddot{v}_0^{m,n} = C_1\left(\Delta_{yy}v_1^{m,n} + 2v_1^{m,n} - 2v_0^{m,n}\right) + \frac{1}{2}C_2\left(\Delta_{xy}u_0^{m,n} + \Delta v_0^{m,n}\right) - AC_1\Delta_y\varphi_1^{m,n} + f_0^y, \quad (4)$$

$$I\ddot{\varphi}_0^{m,n} = A^2C_1\left(\Delta_{xx}\varphi_1^{m,n} + \Delta_{yy}\varphi_1^{m,n} + 4\varphi_1^{m,n} - 4\varphi_0^{m,n}\right) - AC_1\left(\Delta_x u_1^{m,n} + \Delta_y v_1^{m,n}\right) - c\varphi_0^{m,n} + f_0^\varphi.$$

Here we have introduced the following notations for finite differences

$$\Delta_x u_s^{m,n} = u_s^{m+1,n} - u_s^{m-1,n}, \quad \Delta_{xx}u_s^{m,n} = u_s^{m+1,n} - 2u_s^{m,n} + u_s^{m-1,n},$$

$$\Delta_y u_s^{m,n} = u_s^{m,n+1} - u_s^{m,n-1}, \quad \Delta_{yy}u_s^{m,n} = u_s^{m,n+1} - 2u_s^{m,n} + u_s^{m,n-1},$$

$$\Delta_{xy}u_s^{m,n} = u_s^{m+1,n+1} - u_s^{m-1,n+1} - u_s^{m+1,n-1} + u_s^{m-1,n-1}, \quad (5)$$

$$\Delta u_s^{m,n} = u_s^{m+1,n+1} + u_s^{m-1,n+1} + u_s^{m+1,n-1} + u_s^{m-1,n-1} - 4u_s^{m,n}.$$

Three equations for particles with $s=1$ can be obtained from Eq. (4) substituting $0 \to 1$, $1 \to 0$, $m \to m+1$ and $A \to -A$.

Dispersion relations for free vibrations, $f_s^x = f_s^y = f_s^\varphi = 0$, for the discrete model with one particle in a periodic cell have been derived in Vasiliev et al. (2002). In order to derive similar dispersion relations for the discrete system with two particles in a periodic cell we consider solution of the form

$$u_s^{m,n}(t) = \tilde{u}_s \exp[i(\omega t - mK_x - nK_y)],$$

$$v_s^{m,n}(t) = \tilde{v}_s \exp[i(\omega t - mK_x - nK_y)], \quad (6)$$

$$\varphi_s^{m,n}(t) = \tilde{\varphi}_s \exp[i(\omega t - mK_x - nK_y)],$$

where $K_x = hk_x$, $K_y = hk_y$, with $k_x$, $k_y$ being the wave numbers, $\omega$ is the angular frequency, and $\tilde{u}_s, \tilde{v}_s, \tilde{\varphi}_s$ are the amplitudes.

Substituting Eq. (6) into Eq. (4) and similar equations for $s=1$, we obtain six linear equations for amplitudes. In terms of new variables,

$$\tilde{U}_0 = \frac{1}{2}(\tilde{u}_1 + \tilde{u}_0), \quad \tilde{V}_0 = \frac{1}{2}(\tilde{v}_1 + \tilde{v}_0), \quad \tilde{\Phi}_0 = \frac{1}{2}(\tilde{\varphi}_1 - \tilde{\varphi}_0),$$

$$\widetilde{U}_1 = \frac{1}{2}(\widetilde{u}_1 - \widetilde{u}_0), \quad \widetilde{V}_1 = \frac{1}{2}(\widetilde{v}_1 - \widetilde{v}_0), \quad \widetilde{\Phi}_1 = \frac{1}{2}(\widetilde{\varphi}_1 + \widetilde{\varphi}_0), \tag{7}$$

the six equations split into two independent sets of equations of the form,

$$\begin{aligned}
(a_{s,0} + a_{s,2} + m\omega_s^2)\widetilde{U}_s + a_{s,3}\widetilde{V}_s + a_{s,4}\,i\widetilde{\Phi}_s &= 0, \\
a_{s,3}\widetilde{U}_s + (a_{s,1} + a_{s,2} + m\omega_s^2)\widetilde{V}_s + a_{s,5}\,i\widetilde{\Phi}_s &= 0, \\
a_{s,4}\widetilde{U}_s + a_{s,5}\widetilde{V}_s + (a_{s,6} + I\omega_s^2)\,i\widetilde{\Phi}_s &= 0,
\end{aligned} \tag{8}$$

where $s = 0,1$.

For $s = 0$ the coefficients in Eq. (8) are

$$\begin{aligned}
a_{0,0} &= 2C_1(\cos K_x - 1), \quad a_{0,1} = 2C_1(\cos K_y - 1), \quad a_{0,2} = 2C_2(\cos K_x \cos K_y - 1), \\
a_{0,3} &= -2C_2 \sin K_x \sin K_y, \quad a_{0,4} = 2AC_1 \sin K_x, \quad a_{0,5} = 2AC_1 \sin K_y, \\
a_{0,6} &= 2A^2 C_1(-\cos K_x - \cos K_y - 2) - c,
\end{aligned} \tag{9}$$

which coincides with the coefficients derived in Vasiliev et al. (2002) for the case of one particle in a periodic cell.

For $s = 1$ the coefficients are

$$\begin{aligned}
a_{1,0} &= 2C_1(-\cos K_x - 1), \quad a_{1,1} = 2C_1(-\cos K_y - 1), \quad a_{1,2} = 2C_2(\cos K_x \cos K_y - 1), \\
a_{1,3} &= -2C_2 \sin K_x \sin K_y, \quad a_{1,4} = -2AC_1 \sin K_x, \quad a_{1,5} = -2AC_1 \sin K_y, \\
a_{1,6} &= 2A^2 C_1(\cos K_x + \cos K_y - 2) - c.
\end{aligned} \tag{10}$$

Equalizing the determinant of Eq. (8) with $s = 0$ and $s = 1$ to zero one obtains the six branches of the dispersion relations. Substitutions $K_x \to \pi - K_y$ and $K_y \to \pi - K_x$ in the coefficients $a_{1,m}$ given by Eq. (10) lead to the same determinant for dispersion relations as for the coefficients $a_{0,m}$ given by Eq. (9). Thus, the dispersion relations derived for two particles in a periodic cell define six surfaces in the reduced first Brillouin zone, $\|K_x\| + \|K_y\| < \pi$. Three of them, having coefficients $a_{0,m}$, coincide with the surfaces for discrete model with one particle in a periodic cell derived in Vasiliev et al. (2002). Three other, having coefficients $a_{1,m}$, are the same surfaces folded with respect to the four planes $K_x \pm K_y = \pm\pi$ from the region $\|K_x\| + \|K_y\| > \pi$, $|K_x| < \pi$, $|K_y| < \pi$.

## 4. Two-field theory

We introduce two vector-functions $\{u_s(x,y,t), v_s(x,y,t), \varphi_s(x,y,t)\}$, $s = 0,1$, in order to describe the displacements of the particles marked by the indices 0 and 1.

We assume that $\{u_s^{m,n}(t), v_s^{m,n}(t), \varphi_s^{m,n}(t)\} = \{u_s(x,y,t), v_s(x,y,t), \varphi_s(x,y,t)\}\big|_{x=mh,\ y=nh}$, $s = 0,1$.

Substituting in Eq. (4) discrete values with

$$w_s^{m\pm 1, n\pm 1}(t) \to w_s(x \pm h, y \pm h, t) = \sum_i \sum_j \frac{(\pm h)^i}{i!} \frac{(\pm h)^j}{j!} \frac{\partial^{i+j} w_s(x,y,t)}{\partial x^i \partial x^j}, \tag{11}$$

with the use of the notations Eq. (5), we obtain a high-gradient multi-field theory. Assuming that the displacement in the discrete model vary slowly from one macro-cell to another and using the Taylor series expansions, Eq.(11), for field functions up to second order terms, we come to the six equations of the two-field long-wave theory. The first three equations are

$$Mu_{0,tt} = C_1\left(h^2 u_{1,xx} + 2u_1 - 2u_0\right) + C_2 h^2 \left(\Delta u_0 + 2v_{0,xy}\right) - 2AC_1 h\varphi_{1,x} + f_0^x,$$
$$Mv_{0,tt} = C_1\left(h^2 v_{1,yy} + 2v_1 - 2v_0\right) + C_2 h^2 \left(2u_{0,xy} + \Delta v_0\right) - 2AC_1 h\varphi_{1,y} + f_0^y, \quad (12)$$
$$I\varphi_{0,tt} = A^2 C_1\left(h^2 \Delta\varphi_1 + 4\varphi_1 - 4\varphi_0\right) - 2AC_1 h\left(u_{1,x} + v_{1,y}\right) - c\varphi_0 + f_0^\varphi,$$

where subscript indices after commas denote the partial derivatives with respect to coordinates and the notation $\Delta w \equiv w_{xx} + w_{yy}$ was used. Three other equations can be obtained substituting $0 \to 1$, $1 \to 0$, $A \to -A$.

In terms of the following new variables and corresponding combinations of the external forces and torques,

$$U_s = \frac{1}{2}\left[u_1 + (-1)^s u_0\right], \quad V_s = \frac{1}{2}\left[v_1 + (-1)^s v_0\right], \quad \Phi_s = \frac{1}{2}\left[\varphi_1 - (-1)^s \varphi_0\right],$$
$$F_s^x = \frac{1}{2}\left[f_1^x + (-1)^s f_0^x\right], \quad F_s^y = \frac{1}{2}\left[f_1^y + (-1)^s f_0^y\right], \quad F_s^\varphi = \frac{1}{2}\left[f_1^\varphi - (-1)^s f_0^\varphi\right], \quad (13)$$

the six coupled equations of the continuum theory split into two independent groups. Equations for functions $U_0$, $V_0$, $\Phi_0$ are

$$MU_{0,tt} = C_1 h^2 U_{0,xx} + C_2 h^2 \left(\Delta U_0 + 2V_{0,xy}\right) - 2AC_1 h\Phi_{0,x} + F_0^x,$$
$$MV_{0,tt} = C_1 h^2 V_{0,yy} + C_2 h^2 \left(2U_{0,xy} + \Delta V_0\right) - 2AC_1 h\Phi_{0,y} + F_0^y, \quad (14)$$
$$I\Phi_{0,tt} = 2AC_1 h\left(U_{0,x} + V_{0,y}\right) - A^2 C_1\left(h^2 \Delta\Phi_0 + 8\Phi_0\right) - c\Phi_0 + F_0^\varphi.$$

These equations coincide with single-field equations obtained in Vasiliev et al. (2002). However, in the present case we have three more equations for the functions $U_1$, $V_1$, $\Phi_1$,

$$MU_{1,tt} = -C_1\left(h^2 U_{1,xx} + 4U_1\right) + C_2 h^2\left(\Delta U_1 + 2V_{1,xy}\right) + 2AC_1 h\Phi_{1,x} + F_1^x,$$
$$MV_{1,tt} = -C_1\left(h^2 V_{1,yy} + 4V_1\right) + C_2 h^2\left(2U_{1,xy} + \Delta V_1\right) + 2AC_1 h\Phi_{1,y} + F_1^y, \quad (15)$$
$$I\Phi_{1,tt} = -2AC_1 h\left(U_{1,x} + V_{1,y}\right) + A^2 C_1 h^2 \Delta\Phi_1 - c\Phi_1 + F_1^\varphi.$$

Dispersion relations for the continuum theory can be found substituting

$$U_s(x,y,t) = \tilde{U}_s \exp\left[i(\omega t - k_x x - k_y y)\right],$$
$$V_s(x,y,t) = \tilde{V}_s \exp\left[i(\omega t - k_x x - k_y y)\right], \quad (16)$$
$$\Phi_s(x,y,t) = \tilde{\Phi}_s \exp\left[i(\omega t - k_x x - k_y y)\right]$$

into Eq. (14) and Eq. (15) for free vibrations, $F_s^x = F_s^y = F_s^\varphi = 0$. This gives the set of linear algebraic equations of the form of Eq. (8) with coefficients

$$c_{0,0} = -C_1 K_x^2, \quad c_{0,1} = -C_1 K_y^2, \quad c_{0,2} = C_2\left(-K_x^2 - K_y^2\right), \quad c_{0,3} = -2C_2 K_x K_y,$$
$$c_{0,4} = 2AC_1 K_x, \quad c_{0,5} = 2AC_1 K_y, \quad c_{0,6} = A^2 C_1\left(K_x^2 + K_y^2 - 8\right) - c, \quad (17)$$

and

$$c_{1,0} = C_1\left(K_x^2 - 4\right), \quad c_{1,1} = C_1\left(K_y^2 - 4\right), \quad c_{1,2} = C_2\left(-K_x^2 - K_y^2\right), \quad c_{1,3} = -2C_2 K_x K_y,$$
$$c_{1,4} = -2AC_1 K_x, \quad c_{1,5} = -2AC_1 K_y, \quad c_{1,6} = A^2 C_1\left(-K_x^2 - K_y^2\right) - c. \quad (18)$$

The coefficients $c_{0,m}$ are the Taylor series expansions of the coefficients $a_{0,m}$ for discrete model, Eq. (9), in the vicinity of the point $(K_x, K_y) = (0, 0)$. The coefficients $c_{1,m}$ are nothing but second order Taylor expansions of the coefficients $a_{1,m}$, Eq. (10), in the vicinity of the point $(K_x, K_y) = (0, 0)$ and, in accordance to the statement that was made in the conclusion of Sec. 3, corresponding dispersion surfaces folded with respect to the planes $K_x \pm K_y = \pm \pi$ approximate the dispersion surfaces of the discrete model in the vicinity of the points $(K_x, K_y) = (\pm \pi, \pm \pi)$.

Thus, the two-field theory contains three equations (14) of the single-field micropolar theory, Eq. (14), which describes the long-wavelength solutions and three more equations, Eq. (15), describing short-wave solutions with the wave numbers near the corners of the first Brillouin zone, $(\pm \pi, \pm \pi)$. The two-field theory gives the largest error near the points $(\pm \pi, 0)$ and $(0, \pm \pi)$ of the original first Brillouin zone. In Sec. 5 we derive the four-field theory which improves the description of the vibration spectrum in the vicinity of these points.

## 5. Four-field theory

We consider the periodic cell containing four particles with coordinates $(m, n)$, $(m+1, n)$, $(m+1, n+1)$, and $(m, n+1)$ (Fig. 1(a)). Particles with $(m, n) = (2i, 2j)$, $(m, n) = (2i+1, 2j)$, $(m, n) = (2i+1, 2j+1)$, and $(m, n) = (2i, 2j+1)$ have indices $s = 0, 1, 2,$ and 3, respectively.

We construct the discrete Lagrangian Eq. (2) for four particles of the periodic cell and derive twelve equations of motion having structure similar to Eq. (4).

In order to derive the four-field theory we introduce four vector-functions $\{u_s(x,y,t), v_s(x,y,t), \varphi_s(x,y,t)\}$, $s = 0,...,3$. Then, by using Taylor series expansions, we come to the twelve coupled equations of four-field theory in the same way as it was described in Sec. 4 for the two-field theory.

We introduce new variables

$$U_0 = \frac{1}{4}(u_0 + u_1 + u_2 + u_3), \quad U_1 = \frac{1}{4}(u_1 - u_0 + u_3 - u_2),$$
$$U_2 = \frac{1}{4}(u_1 - u_0 + u_2 - u_3), \quad U_3 = \frac{1}{4}(u_3 - u_0 + u_2 - u_1) \quad (19)$$

and similar combinations $V_s$ of displacements $v_s$ and combinations $F_s^x$ ($F_s^y$) of forces $f_s^x$ ($f_s^y$). We also introduce the following new variables for the rotational degrees of freedom

$$\Phi_0 = \frac{1}{4}(\varphi_1 - \varphi_0 + \varphi_3 - \varphi_2), \quad \Phi_1 = \frac{1}{4}(\varphi_0 + \varphi_1 + \varphi_2 + \varphi_3),$$
$$\Phi_2 = \frac{1}{4}(\varphi_3 - \varphi_0 + \varphi_2 - \varphi_1), \quad \Phi_3 = \frac{1}{4}(\varphi_1 - \varphi_0 + \varphi_2 - \varphi_3) \quad (20)$$

and similar combinations $F_s^\varphi$ of torques $f_s^\varphi$.

In terms of new variables the twelve equations of motion split into four independent groups. For $U_0$, $V_0$, $\Phi_0$, equations coincide with Eq. (14) and for $U_1$, $V_1$, $\Phi_1$, they coincide with Eq. (15).

We have six new equations, for $U_2$, $V_2$, $\Phi_2$,

$$MU_{2,tt} = -C_1(h^2 U_{2,xx} + 4U_2) - C_2[h^2(\Delta U_2 + 2V_{2,xy}) + 4U_2] - 2AC_1 h \Phi_{2,x} + F_2^x,$$
$$MV_{2,tt} = C_1 h^2 V_{2,yy} - C_2[h^2(2U_{2,xy} + \Delta V_2) + 4V_2] + 2AC_1 h \Phi_{2,y} + F_2^y, \quad (21)$$
$$I\Phi_{2,tt} = 2AC_1 h(U_{2,x} - V_{2,y}) + A^2 C_1[h^2(\Phi_{2,xx} - \Phi_{2,yy}) - 4\Phi_2] - c\Phi_2 + F_2^\varphi,$$

and $U_3$, $V_3$, $\Phi_3$,

$$MU_{3,tt} = C_1 h^2 U_{3,xx} - C_2 \left[ h^2 \left( \Delta U_3 + 2V_{3,xy} \right) + 4U_3 \right] + 2AC_1 h \Phi_{3,x} + F_3^x,$$

$$MV_{3,tt} = -C_1 \left( h^2 V_{3,yy} + 4V_3 \right) - C_2 \left[ h^2 \left( 2U_{3,xy} + \Delta V_3 \right) + 4V_3 \right] - 2AC_1 h \Phi_{3,y} + F_3^y, \quad (22)$$

$$I\Phi_{3,tt} = 2AC_1 h \left( -U_{3,x} + V_{3,y} \right) + A^2 C_1 \left[ h^2 \left( -\Phi_{3,xx} + \Phi_{3,yy} \right) - 4\Phi_3 \right] - c\Phi_3 + F_3^\varphi.$$

Substitution of Eq. (16) with $s = 0,...,3$ into Eqs. (14), (15), (21), and (22) gives the possibility to obtain the dispersion relations from four independent sets of three linear algebraic equations. Coefficients of the sets for $s = 0$ and $s = 1$ coincide with that given by Eq. (17) and Eq. (18), respectively.

For $s = 2$ coefficients read

$$c_{2,0} = C_1 \left( K_x^2 - 4 \right), \quad c_{2,1} = -C_1 K_y^2, \quad c_{2,2} = C_2 \left( K_x^2 + K_y^2 - 4 \right), \quad c_{2,3} = 2C_2 K_x K_y,$$

$$c_{2,4} = 2AC_1 K_x, \quad c_{2,5} = -2AC_1 K_y, \quad c_{2,6} = A^2 C_1 \left( -K_x^2 + K_y^2 - 4 \right) - c \quad (23)$$

and for $s = 3$ they read

$$c_{3,0} = -C_1 K_x^2, \quad c_{3,1} = C_1 \left( K_y^2 - 4 \right), \quad c_{3,2} = C_2 \left( K_x^2 + K_y^2 - 4 \right), \quad c_{3,3} = 2C_2 K_x K_y,$$

$$c_{3,4} = -2AC_1 K_x, \quad c_{3,5} = 2AC_1 K_y, \quad c_{3,6} = A^2 C_1 \left( K_x^2 - K_y^2 - 4 \right) - c. \quad (24)$$

Dispersion surfaces corresponding to the coefficients $c_{2,m}$ folded with respect to the planes $K_x = \pm \pi / 2$ coincide with the Taylor expansions of the dispersion surfaces of the discrete model in the vicinity of the points $(K_x, K_y) = (\pm \pi, 0)$. Dispersion surfaces corresponding to the coefficients $c_{3,m}$ folded with respect to the planes $K_y = \pm \pi / 2$ coincide with Taylor expansions of the dispersion surfaces of the discrete model in the vicinity of the points $(K_x, K_y) = (0, \pm \pi)$.

Thus, the four-field continuum theory contains the equations of two-field theory, Eqs. (14), (15), and improves the two-field theory in the vicinity of the points $(K_x, K_y) = (0, \pm \pi)$ and $(K_x, K_y) = (\pm \pi, 0)$, i.e., it is capable of description of the short waves with the wave vectors close to these points.

## 6. Multi-field modeling of localized distortions in crystal

In this section, multi-field modeling of static localized distortions in the 2D model of crystal is considered. We obtain a general solution and solve a boundary value problem for the discrete system and then derive a multi-field continuum approximation for the obtained solutions.

We consider one-dimensional problem assuming that displacements of particles do not change with $n$,

$$u_s^{m,n} = u_m, \quad v_s^{m,n} = v_m, \quad \varphi_s^{m,n} = (-1)^{m+n} \varphi_m \quad (25)$$

and simplify the equations of motion Eq. (4):

$$M\ddot{u}_m = (C_2 + C_1)(u_{m+1} - 2u_m + u_{m-1}) + AC_1 (\varphi_{m+1} - \varphi_{m-1}),$$

$$I\ddot{\varphi}_m = -AC_1 (u_{m+1} - u_{m-1}) - A^2 C_1 (\varphi_{m+1} + 6\varphi_m + \varphi_{m-1}) - c\varphi_m. \quad (26)$$

With the use of the static solution to Eq. (26) one can write the solution in terms of variables Eq. (25) for even $n$ in the form

$$u_s^m = \alpha_0 + mh\alpha_1 + (-1)^s e^{-\lambda mh} \alpha_2 + (-1)^s e^{\lambda mh} \alpha_3,$$

$$\varphi_s^m = (-1)^s r_1 \alpha_1 + r_2 e^{-\lambda mh} \alpha_2 - r_2 e^{\lambda mh} \alpha_3, \quad (27)$$

where $s = 0, 1$, and

$$r_1 = -\frac{2AC_1 h}{8A^2 C_1 + c}, \quad r_2 = \frac{C_1 + C_2}{AC_1} \frac{2[1 + \cosh(\lambda h)]}{2\sinh(\lambda h)}. \tag{28}$$

The inverse width of the localized distortion, $\lambda > 0$, is defined in terms of the model parameters as

$$4\sinh^2(\lambda h) = (1 + C_1/C_2)(4 + c/C_1 A^2), \tag{29}$$

which is obtained from the characteristic equations for Eq. (26) in static case

$$\det \begin{bmatrix} 2(C_1 + C_2)[\cosh(\lambda h) + 1] & 2AC_1 \sinh(\lambda h) \\ 2AC_1 \sinh(\lambda h) & 2A^2 C_1 [\cosh(\lambda h) - 3] - c \end{bmatrix} = 0. \tag{30}$$

Constants $\alpha_j$, $j = 0, 1, 2, 3$ in Eq. (27) must be chosen to satisfy boundary conditions.

For example, for the crystal layer, $0 < m < N$, solution to Eq. (27) satisfying the boundary conditions

$$u_0^0 = 0, \quad \varphi_0^0 = 0, \quad u_1^N = u_N, \quad \varphi_1^N = 0, \tag{31}$$

has the form

$$u_s^m = \left[1 + r_2 mh/r_1 - (-1)^s e^{-\lambda mh} - (-1)^s e^{-\lambda(N-m)h}\right] p u_N,$$
$$\varphi_s^m = \left[(-1)^s - e^{-\lambda mh} + e^{-\lambda(N-m)h}\right] r_2 p u_N, \tag{32}$$

where $p = 1/(2 + r_2 Nh/r_1)$. For sake of simplicity we have assumed that $N$ is sufficiently large so that the distortions introduced by the boundary conditions at $m = 0$ and $m = N$ do not overlap and we have neglected the term $e^{-\lambda Nh}$ in comparison with 1 in Eq. (32).

Let us now apply the multi-field approach to solve the above problem. We simplify the four-field theory Eqs. (14), (15), (21) and (22) applying the assumptions Eq. (25) which, in view of Eqs. (19) and (20), suggest that the fields $U_s(x,y)$, $V_s(x,y)$, $\Phi_s(x,y)$ are zero for $s = 1$ and $s = 3$.

We start from the single-field theory Eq. (14) which in the one-dimensional case considered here obtains the form

$$MU_{0,tt} = (C_1 + C_2)h^2 U_{0,xx} - 2AC_1 h \Phi_{0,x},$$
$$I\Phi_{0,tt} = 2AC_1 h U_{0,x} - A^2 C_1 (h^2 \Phi_{0,xx} + 8\Phi_0) - c\Phi_0. \tag{33}$$

Characteristic polynomial for the static problem defined by Eq. (33) has a second-order root equal to zero. Corresponding static solution has the form

$$U_0(x) = \alpha_0 + \alpha_1 x,$$
$$\Phi_0(x) = -R_1 \alpha_1, \tag{34}$$

where $R_1 = -2AC_1 h/(8A^2 C_1 + c)$, and $\alpha_0$, $\alpha_1$ are arbitrary constants.

Note that the two other roots of the characteristic polynomial of Eq. (33) are purely imaginary and hence, the general solution Eq. (33) corresponding to the single-field theory does not contain exponentially decaying terms describing the localized distortions.

Exponential part of the solution Eq. (27) can be obtained from Eq. (21) of the four-field theory which, after the reduction to one-dimension, obtains the following form

$$MU_{2,tt} = -(C_2 + C_1)(h^2 U_{2,xx} + 4U_2) - 2AC_1 h \Phi_{2,x},$$
$$I\Phi_{2,tt} = 2AC_1 h U_{2,x} + A^2 C_1 (h^2 \Phi_{2,xx} - 4\Phi_2) - c\Phi_2. \tag{35}$$

Equation (35) has the static solution

$$U_2(x) = -e^{-\Lambda x} \alpha_2 - e^{\Lambda x} \alpha_3,$$
$$\Phi_2(x) = -R_2 e^{-\Lambda x} \alpha_2 + R_2 e^{\Lambda x} \alpha_3, \tag{36}$$

where

$$R_2 = \frac{C_1 + C_2}{AC_1} \frac{4 + (\Lambda h)^2}{2\Lambda h}.$$  (37)

The inverse width of the distortion, $\Lambda > 0$, can be obtained from the characteristic polynomial

$$\det \begin{bmatrix} (C_1 + C_2)[(\Lambda h)^2 + 4] & 2AC_1 \Lambda h \\ 2AC_1 \Lambda h & A^2 C_1[(\Lambda h)^2 - 4] - c \end{bmatrix} = 0.$$  (38)

With the use of the solutions Eqs. (34), (36) and the following relations

$$u_s(x) = U_0(x) - (-1)^s U_1(x),$$
$$\varphi_s(x) = -(-1)^s \Phi_0(x) - \Phi_1(x),$$  (39)

we express the solution in terms of the original variables:

$$u_s(x) = \alpha_0 + \alpha_1 x + (-1)^s e^{-\Lambda x} \alpha_2 + (-1)^s e^{\Lambda x} \alpha_3,$$
$$\varphi_s(x) = (-1)^s R_1 \alpha_1 + R_2 e^{-\Lambda x} \alpha_2 - R_2 e^{\Lambda x} \alpha_3.$$  (40)

Arbitrary constants $\alpha_j$, $j = 0,\ldots,3$, in Eq. (40) are to be found from the boundary conditions. Analog of the boundary conditions Eq. (31) is

$$u_0(0) = 0, \quad \varphi_0(0) = 0, \quad u_1(L) = u_N, \quad \varphi_1(L) = 0,$$  (41)

and corresponding solution has the form

$$u_s(x) = \left[1 + R_2 x / R_1 - (-1)^s e^{-\Lambda x} - (-1)^s e^{-\Lambda(L-x)}\right] P u_N,$$
$$\varphi_s(x) = \left[(-1)^s - e^{-\Lambda x} + e^{-\Lambda(L-x)}\right] R_2 P u_N,$$  (42)

where $P = 1/(2 + R_2 L / R_1)$.

Note that the exact solutions Eq. (27) for the discrete system and corresponding multi-field solution, Eq. (40), have similar structures. Equation (38) for the parameter $\Lambda$ of multi-field theory can be obtained from Eq. (30) for the corresponding parameter $\lambda$ expanding the hyperbolic functions in Taylor series up to second order terms. The formulae for parameters $R_1$, $R_2$, $P$ of approximate continuum solution can also be obtained by Taylor expansion of the hyperbolic functions in the formulae for corresponding parameters $r_1$, $r_2$, $p$ in the discrete solution. The differences between the hyperbolic functions and corresponding Taylor series expansions up to second order around the zero point define the accuracy of the continuum solutions.

Further analysis and comparison of the discrete model Eq. (26) with the multi-field model Eqs. (33), (35) will be carried out with respect to the dynamical solutions of the form

$$u_m(t) = \tilde{u} \exp[i\omega t - mK],$$
$$\varphi_m(t) = \tilde{\varphi} \exp[i\omega t - mK],$$  (43)

for the discrete system and of the form

$$U_s(x,t) = \tilde{u}_s \exp[i\omega t - Kx/h],$$
$$\Phi_s(x,t) = \tilde{\varphi}_s \exp[i\omega t - Kx/h],$$  (44)

for the multi-field continuum, where $K$ is a complex parameter.

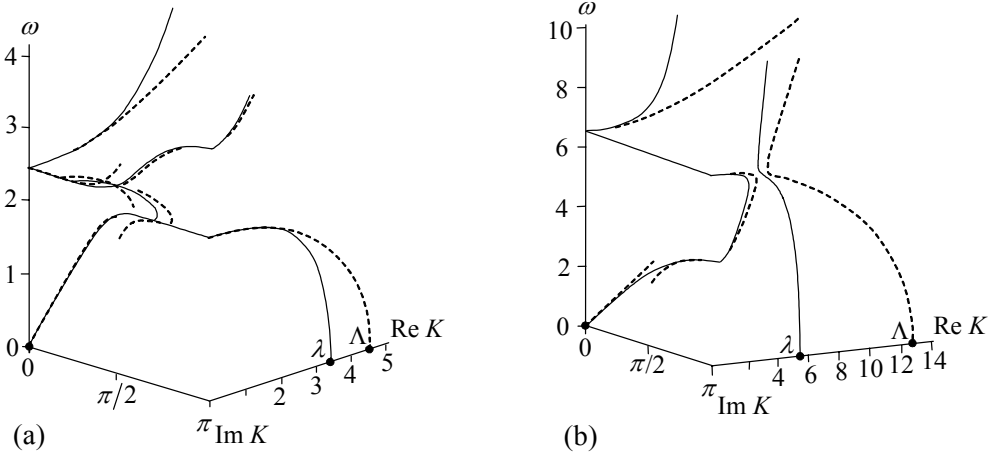

Fig.2. Relations between $\omega$ and $\text{Re}\,K$, $\text{Im}\,K$ for the solutions of the form Eq.(43) and Eq. (44). Solid lines show the exact dispersion curves for the discrete model. The dispersion curves for the multi-field theory are presented by dashed lines. Dots on the $\omega = 0$ plane show the roots of characteristic equations for static problem. In (a) and (b) we contrast the results obtained for comparatively weak and rather strong rotational background potential setting $c = 4$ and $c = 40$, respectively. Other parameters are $C_1 = 1$, $C_2 = 2$, $A = 0.5$, $h = 1$, $M = 1$, $I = 1$.

Substituting Eq. (43) and (44) into corresponding equations of motion we obtain the relations between $\omega$ and $\text{Re}\,K$, $\text{Im}\,K$. We compare these relations for the discrete model and multi-field theory in Fig. 2. Without the loss in generality, one can fix, for example, $C_1$, $h$, and $M$. Then we introduce the dimensionless quantities $\bar{u}_m = u_m / h$, $\bar{t} = t / \sqrt{C_1 / M}$, $\bar{x} = x / h$, set $C_2 = 2C_1$, $A = 0.5h$, $I = Mh^2$, and, as in Vasiliev et al. (2002), consider two different magnitudes of rotational background potential, $c = 4h^2 C_1$ and $c = 40h^2 C_1$, in Fig.2(a) and Fig. 2(b), respectively. Curves for the discrete system are shown by solid lines and for multi-field theory by dashed lines. The roots of characteristic equations in static case, i.e., in the $\omega = 0$ plane, are depicted by dots. $K = 0$ is a second-order root for both discrete model and continuum theory. Two other roots differ noticeably for discrete system and multi-field theory, but qualitatively they belong to similar branches in the plane $\text{Im}\,K = \pi$. Solid and dashed curves are tangent at the point $K = (\pi, 0)$. The difference is small for $K$ near this point and increases with the distance from it. Difference in the curves for the discrete and continuum theories is rather large in the static case $(\omega = 0)$ at the considered set of parameters. This can be explained by the very sharp localization of the distortion in our examples. However, even for such a sharp localization which cannot be accurately captured by a long-wave continuum theory we have obtained a very good qualitative agreement between discrete and multi-field theories because the multi-field theory contains short-wave decaying solution which does not exist in the single-field theory. As it can be seen from Eq. (29), the width of the localized distortion can be very large if at least one of the model parameters, $C_1$, $C_2$, or $c$, can obtain negative values. In this case, the multi-field theory will give not only qualitative but also quantitative solutions.

As an illustration, in Fig. 3 we present the values $\bar{\varphi}_s^m = \varphi_s^m / |r_2 p u_N|$ given by Eq. (32) by dots and plot the dashed zigzag line to show the short-wave character of the solution. Two solid lines show the two-field solution $\bar{\varphi}_s(x) = \varphi_s(x) / |r_2 p u_N|$, Eq. (42), which approximately describes the rapidly oscillating solution in the discrete system, one solid line for the points $2mh$ and another one for the points $(2m+1)h$. Near the points $m = 0$ and $m = N$, the zigzag solution $(-1)^s r_2 p u_N$ is corrected by the localised solutions in order to satisfy the boundary conditions.

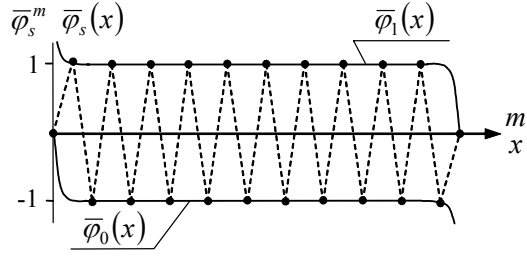

Fig. 3. Comparison of the rotational displacements for the boundary-value problem considered in Sec. 6. Dots show the exact solution for the discrete model, $\bar{\varphi}_s^m$, defined by Eq. (32). The dashed zigzag line is plotted to illustrate the short-wave character of the solution. Two smooth functions, $\bar{\varphi}_0(x)$ and $\bar{\varphi}_1(x)$, represent the multi-field solution, Eq. (42), for odd and even particles, respectively. Parameters of the model are set as in Fig. 2a.

Multi-field solution Eq. (32) describes not only slowly varying displacements $u_m^s$ in the discrete system (the first two terms in the first equations of Eq. (32)) but also the short-wave spatially localized displacements (the third and fourth terms). The latter are not reproduced by the single-field micropolar model Eqs. (14), (33).

## 7. Discussion

In this work, for the 2D discrete model with finite size particles, we have constructed two- and four-field continuum theories. Standard single-field continuum theory derived in Vasiliev et al. (2002) was valid only for wave numbers close to the origin of the first Brillouin zone while the two-field theory improves the approximation of the discrete spectrum for the short waves near the points $(\pm\pi,\pm\pi)$ and the four-field theory gives an additional improvement near the points $(\pm\pi,0)$ and $(0,\pm\pi)$. Consideration of a periodic cell containing several primitive cells leads to the reduction of area of the first Brillouin zone and folding of the dispersion surfaces of discrete system. For two particles in a periodic cell the folding occurs with respect to planes $K_x \pm K_y = \pm\pi$ and points $(K_x, K_y) = (\pm\pi, \pm\pi)$ become $(0,0)$ point in the new reciprocal basis. The two-field theory is accurate at these points. Dispersion surfaces for the cell with four particles can be obtained by folding the original dispersion surfaces with respect to $K_x = \pm\pi/2$ and $K_y = \pm\pi/2$. As a result of folding, points $(K_x, K_y) = (\pm\pi, \pm\pi)$, and also $(K_x, K_y) = (0, \pm\pi)$, and $(K_x, K_y) = (\pm\pi, 0)$ become $(0,0)$ points and the four-field continuum theory is valid for short waves in the vicinity of these points.

Generally speaking, the $N$-field theory is constructed as a continuum analogue for the discrete system with a periodic cell containing $N$ primitive cells. Consideration of more than one primitive cell in a periodic cell is meaningless for a discrete periodic system because the dispersion surfaces in this case are just the original dispersion surfaces $N$ times folded in the $N$ times reduced first Brillouin zone. The number of equations of motion increases by $N$ times but they do not contain any new information. However, the $N$-field continuum theory derived for a periodic cell with $N$ primitive cells gives a piecewise approximation to the exact dispersion surface and the number of pieces is equal to $N$. Each piece approximates the exact dispersion surface in the vicinity of point $(K_x, K_y) = (0,0)$ of the reduced reciprocal basis. Thus, the $N$-field theory gives a good approximation of dispersion surfaces not only for long waves but also for short waves and, in the case of multiple folding, for the waves inside the first Brillouin zone.

In Fig. 4 we compare the exact dispersion curves of the discrete model (solid lines) with the single-field continuum theory derived in Vasiliev et al. (2002) (dotted lines) and its two different generalizations, namely, the high-gradient (dashed lines in (a)) and the multi-field (dashed lines in (b)) theories. The single-field theory gives a good approximation only near the origin of the first Brillouin zone. The high-gradient theory (fourth order terms retained in Taylor expansions) extends the region of validity of the long-wave theory but gives no improvement near the zone boundary, $K_x = \pi$. In (b), the three branches of the dispersion curves for the four-field continuum theory, Eqs. (14), (21), are shown unfolded. Near the origin the multi-field theory gives the accuracy of the single-field theory (second order). The four-field theory also gives a good approximation near the zone boundary, and it is exact at the point $K_x = \pi$, where the single-field and the high-gradient theories give maximum error. Thus, the high-gradient and the multi-field approaches improve the long-wave single-field theory in different ways. Taking into account the higher order terms in the Taylor series expansions Eq. (11) the high-gradient multi-field theories can be constructed so that one can combine the advantages of both theories.

The present study is based on the assumption of rigid finite size particles, which is often used in the solid state physics and materials science (Grima and Evans, 2000; Ishibashi and Iwata, 2000; Swainson and Dove, 1993; Dove et al., 1997; Wells et al., 2002; Dmitriev et al., 1997, 1998, 2000, 2003b; Vasiliev et al., 2002). If necessary, the model can be generalized for the case of deformable particles taking into account additional degrees of freedom.

Since the multi-field theory is valid for both long and short waves, it is an appropriate theory to describe the coupling between them. The physical situation where short and long waves are strongly coupled can be easily realized in periodic systems (some natural crystals or manmade structures) having comparatively rigid finite size particles with rotational degrees of freedom. This can be intuitively understood, for example, by inspection of the structure presented in Fig. 1. It can be seen that a homogeneous deformation, e.g., uniaxial or hydrostatic compression, results in staggered rotations of particles. Such coupling of long and short waves usually does not take place for materials consisting of pointwise particles or for materials having finite size particles linked such that the mutual rotations are suppressed or they are not of a staggered type in response to a homogeneous strain. For the materials of these kinds the use of well developed classical or single-field micropolar theories can be sufficient. However, when short and long waves are strongly coupled the use of a multi-field theory can be indispensable. Some of the crystalline materials having finite size particles with rotational degrees of freedom, for which the coupling between long and short waves can be important, we would like to mention the ones described in the Introduction, i.e., polymorphs of silica ($SiO_2$), $KH_2PO_4$ (KDP) family of crystals, some perovskites, e.g., $SrTiO_3$. It has been also proved that microscopic rotations can be responsible for the negative Poisson ratio exhibited by some natural and manmade auxetic materials (Grima and Evans, 2000; Ishibashi and Iwata, 2000; Vasiliev et al., 2002; Wojciechowski et. al., 2003; Wojciechowski 2003).

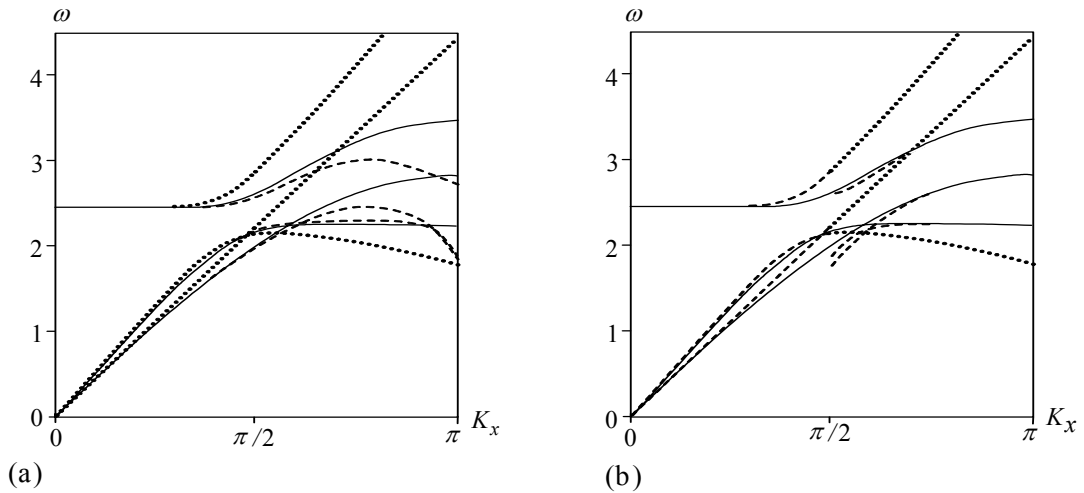

Fig. 4. Comparison of the exact dispersion surfaces for the discrete model (solid lines) with that for different continuum theories (section $K_y = 0$). Dispersion relations of the long-wave approximation (single-field theory) derived in Vasiliev et al. (2002) are plotted in (a) and (b) by dotted lines. Dashed lines show (a) the single-field high-gradient theory (fourth order terms retained in Taylor expansions) and (b) the four-field continuum theory. Parameters of the model are set as in Fig. 2a.

It is important to note that the multi-field approach gives a natural way of incorporating the short-wave external fields of forces and torques into a continuum theory. For example, for the model studied in this paper, the short-wave external force (or torque) of the form $(-1)^{m+n} f$, transforms to a smooth field of forces, $F_0 = 0$, $F_1 = -f$, $F_2 = 0$, $F_3 = 0$ (or torques, $F_0 = -f$, $F_1 = 0$, $F_2 = 0$, $F_3 = 0$), and they can be taken into account in frame of the four-field continuum theory, Eqs. (14), (15), (21), (22).

## 8. Conclusions

For the 2D discrete model of KDP crystal we have derived the two- and four-field micropolar continuum theories. The $N$-field theory is obtained as a continuum analogue for the discrete model with a periodic cell containing $N$ primitive cells by using $N$ vector fields to describe deformation of the crystal. This approach gives the possibility to construct a hierarchy of theories with increasing complexity and accuracy.

Comparison of the dispersion relations for the multi-field and the high-gradient theories suggests that the latter improves the classical long-wave theory in the vicinity of small wave numbers while the former near the zone boundaries and, if desired, inside the first Brillouin zone. We have shown that the multi-field approach makes it possible to describe the localised structural distortions rapidly oscillating in space.

The use of several fields describing external forces and/or torques makes it possible to construct a continuum theory taking into account the external fields rapidly changing in space.

Since the multi-field theory is valid for both long and short waves, it is an appropriate theory to describe the coupling between them, mode softening, and other related phenomena for the crystals with the rotational degrees of freedom. The multi-field theory can be also useful in the rapidly developing theory of the lattice instability (VanVliet et al., 2003; Dmitriev et al., 2004a, 2004b), particularly in the study of the post-critical behaviour for the instability with the mode softening inside the first Brillouin zone (modulated and incommensurate phase). One example of the successful use of multi-field approach can be found in Dmitriev et al. (1997) where the domain-wall regime of incommensurate phase has been described.

**Acknowledgments**

Authors are grateful to Y. Ishibashi, T. Shigenari, and K. W. Wojciechowski for the fruitful discussions.

**References**

Aifantis, E.C., 2003. Update on a class of gradient theories. Mechanics of Materials 35, 259-280.
Alderson, A., Evans, K.E., 2002. Molecular origin of auxetic behavior in tetrahedral framework silicates. Phys. Rev. Lett. 89, 225503.
Askes, H., Suiker, A.S.J., Sluys, L.J., 2002. A classification of higher-order strain-gradient models – linear analysis. Arch. Appl. Mech. 72, 171-188.
Braun, O.M., Kivshar, Y.S., 2004. The Frenkel-Kontorova Model: Concepts, Methods, and Applications. Springer, Berlin.
Cosserat, E., Cosserat, F., 1909. Théorie des Corps Déformables, Hermann A. et Fils, Paris.
Dmitriev, S.V, Shigenari, T., Vasiliev, A.A., Abe, K., 1997. Dynamics of domain walls in an incommensurate phase near the lock-in transition: one-dimensional crystal model. Phys. Rev. B 55, 8155-8164.
Dmitriev, S.V., Shigenari, T., Abe, K., 1998. Mechanisms of transition between 1q and 2q incommensurate phases in two-dimensional crystal model. Phys. Rev. B 58, 2513-2522.
Dmitriev, S.V., Abe, K., Shigenari, T., 2000. Domain wall solutions for EHM model of crystal: structures with period multiple of four. Physica D 147, 122-134.
Dmitriev, S.V., Semagin, D.A., Shigenari, T., Abe, K., Nagamine, M., Aslanyan, T.A., 2003a. Molecular and lattice dynamical study on modulated structures in quartz. Phys. Rev. B 68, 052101.
Dmitriev, S.V., Vasiliev, A.A., Yoshikawa, N., 2003b. Microscopic rotational degrees of freedom in solid state physics. Recent Res. Devel. Physics 4, 267-286.
Dmitriev, S.V., Li, J., Yoshikawa, N., Shibutani, Y., 2004a. Theoretical strength of 2D hexagonal crystals: application to bubble raft indentation. Phil. Mag., (in press).
Dmitriev, S.V., Kitamura, T., Li, J., Umeno, Y., Yashiro, K., Yoshikawa, N., 2004b. Near-surface lattice instability in 2D fiber and half-space. Acta Mater. (in press).

Dove, M.T., Harris, M.J., Hannon, A.C., Parker, J.M., Swainson, I.P., Gambhir, M. 1997. Floppy modes in crystalline and amorphous silicates. Phys. Rev. Lett. 78, 1070-1073.
Eringen, A.C., 1999. Microcontinuum Field Theories: Foundations and Solids. Springer-Verlag, New York.
Fleck, N.A., Hutchinson, J.W., 1997. Strain gradient plasticity. Adv. Appl. Mech. 33, 295-361.
Fleck, N.A., Hutchinson, J.W., 2001. A reformulation of strain gradient plasticity. J. Mech. Phys. Solids 49, 2245-2271.
Forest, S., Pradel, F., Sab, K., 2001. Asymptotic analysis of heterogeneous Cosserat media. Int. J. Solids Struct. 38, 4585-4608.
Grima, J.N., Evans, K.E., 2000. Auxetic behavior from rotating squares. J. Mater. Sci. Lett. 19, 1563-1565.
Il'iushina, E.A., 1969. On a model of continuous medium, taking into account the microstructure. Prikl. Mat. Mekh. 33, 917-923.
Il'iushina, E.A., 1972. A version of the couple stress theory of elasticity for a one-dimensional continuous medium with inhomogeneous periodic structure. Prikl. Mat. Mekh. 36, 1086-1093.
Ishibashi, Y., Iwata, M., 2000. A microscopic model of a negative Poisson's ratio in some crystals. J. Phys. Soc. Jpn. 69, 2702-2703.
Keskar, N.R., Chelikowsky, J.R., 1993. Anomalous elastic behavior in crystalline silica. Phys. Rev. B 48, 16227-16233.
Kimizuka, H., Kaburaki, H., Kogure Y., 2000. Mechanism for negative Poisson ratios over the *alpha-beta* transition of cristobalite, $SiO_2$: a molecular-dynamics study. Phys. Rev. Lett. 84, 5548-5551.
Kuwazuru, O, Yoshikawa, N., 2004a. Theory of elasticity for plain-weave fabrics (1st report, new concept of pseudo-continuum model). JSME Int. J. 47, 17-25.
Kuwazuru, O, Yoshikawa, N., 2004b. Theory of elasticity for plain-weave fabrics (2nd report, finite element formulation). JSME Int. J. 47, 26-34.
Lakes, R.S., 1991. Experimental micro mechanics methods for conventional and negative Poisson's ratio cellular solids as Cosserat continua. J. Engineering Materials and Technology 113, 148-155.
Maugin, G.A., 1999. Nonlinear Waves in Elastic Crystals. Oxford University Press, Oxford.
Mühlhaus, H.-B., Oka, F., 1996. Dispersion and wave propagation in discrete and continuous models for granular materials. Int. J. Solids Struct. 33, 2841-2858.
Noor, A.K., 1988. Continuum modelling for repetitive lattice structures. Appl. Mech. Rev. 41, 7, 285-296.
Peerlings, R.H.J., Geers, M.G.D., de Borst, R., Brekelmans, W.A.M., 2001. A critical comparison of nonlocal and gradient enhanced softening continua. Int. J. Solids Struct. 38, 7723-7746.
Pouget, J., Askar, A., Maugin, G.A., 1986. Lattice model for elastic ferroelectric crystals: continuum approximation. Phys. Rev. B 33, 6320-6325.
Rogula, D. , 1985. Non-classical material continua. Theor. and Appl. Mechanics, Proc. XVIth Int. Congr., 339-353 .
Smirnov, M.B., Mirgorodsky, A.P., 1997. Lattice-dynamical study of the *alpha - beta* phase transition of quartz: soft-mode behavior and elastic anomalies. Phys. Rev. Lett. 78, 2413-2416.
Smirnov, M. B., 1999. Lattice dynamics and thermal expansion of quartz. Phys. Rev. B 59, 4036-4043.
Suiker, A.S.J., de Borst, R., Chang, C.S., 2001. Micro-mechanical modelling of granular material. Part 1: Derivation of a second-gradient micro-polar constitutive theory. Acta Mech. 149, 161-180.
Swainson, I.P., Dove, M.T., 1993. Low-frequency floppy modes in $\beta$ -cristobalite. Phys. Rev. Lett. 71, 193-196.
Vallade, M., Berge, B., Dolino, G., 1992. Origin of the incommensurate phase of quartz: II. Interpretation of inelastic neutron scattering data. J. Physique I 2, 1481-1495.
Van Vliet, K.J., Li, J., Zhu, T., Yip, S., Suresh, S., 2003. Quantifying the early stages of plasticity through nanoscale experiments and simulations. Phys. Rev. B 67, 104105.
Vasiliev, A.A., 1994. The multiple-field approach in modeling stability of a cylindrical shell stiffened with frames at external pressure. Moscow Univ. Mech. Bull. 49(4), 71-74.
Vasiliev, A.A., 1996. Continual modeling of two-row finite discrete system deformation with regard for boundary effects. Moscow Univ. Mech. Bull. 51(5), 44-46.
Vasiliev, A.A., Dmitriev, S.V., Ishibashi, Y., Shigenari, T., 2002. Elastic properties of a two-dimensional model of crystals containing particles with rotational degrees of freedom. Phys. Rev. B 65, 094101.
Wells, S.A., Dove, M.T., Tucker, M.G., Trachenko, K.O., 2002. Real-space rigid unit mode analysis of dynamic disorder in quartz, cristobalite and amorphous silica. J. Phys.: Condens. Matter 14, 4645-4657.
Wojciechowski, K.W., Tretiakov, K.V., Kowalik, M., 2003. Elastic properties of dense solid phases of hard cyclic pentamers and heptamers in two dimensions. Phys. Rev. E 67, 036121.
Wojciechowski, K.W., 2003. Non-chiral, molecular model of negative Poisson ratio in two dimensions. J. Phys. A: Math. Gen. 36, 11765-11778.